\documentclass[twocolumn,showpacs,preprintnumbers,amsmath,amssymb]{revtex4}

\usepackage{graphicx}% Include figure files 
\usepackage{dcolumn}% Align table columns on decimal point
\usepackage{bm}% bold math
\usepackage{epsfig}
\usepackage{epstopdf}
\usepackage{multirow}

\begin{document}

%\preprint{v14.5}
\title{Uncertainties in the Anti-neutrino Production at Nuclear Reactors}

\author{Z. Djurcic$^{1}$}
\email{zdjurcic@nevis.columbia.edu}
\author{J.A. Detwiler$^2$}
\email{JADetwiler@lbl.gov}
\author{A. Piepke$^3$}
\author{V.R. Foster Jr.$^4$}
\author{L. Miller$^5$}
\altaffiliation
[Present address:
]{Sierra Nevada Corp., Los Gatos, CA 95032}
\author{G. Gratta$^5$}
\affiliation{
$^1$Department of Physics, Columbia University, New York, NY 10027
}
\affiliation{
$^2$Lawrence Berkeley National Laboratory, Berkeley, CA 94720
}
\affiliation{
$^3$Department of Physics and Astronomy, University of Alabama, Tuscaloosa, AL 35487
}
\affiliation{
$^4$Diablo Canyon Power Plant, Pacific Gas and Electric Company, Avila Beach, CA 93424
}
\affiliation{
$^5$Physics Department, Stanford University, Stanford, CA 94305
}

\date{\today}

\begin{abstract}
Anti-neutrino emission rates from nuclear reactors are 
determined from thermal power measurements and fission
rate calculations. The uncertainties in these quantities for
commercial power plants and their impact on the calculated 
interaction rates in $\bar{\nu}_{e}$ detectors is examined. 
We discuss reactor-to-reactor correlations between the leading uncertainties,
and their relevance to reactor $\bar{\nu}_e$ experiments. 
\end{abstract}

 % PACS, the Physics and Astronomy Classification Scheme.

\maketitle

\section{Introduction}

Electron anti-neutrinos from large commercial nuclear reactors are playing an important role 
in the exploration of neutrino oscillations~\cite{bemporad}. The choice of distance between source 
and detector allows one to conveniently tune an experiment's sensitivity to either the atmospheric 
or solar neutrino mass splitting.   KamLAND~\cite{kamland_prl,kamland_prl_2,kamland_prl_3}
has established a connection between the MSW effect in the sun~\cite{MSW} and vacuum oscillations
with anti-neutrinos, and has provided the most accurate measurement of $\Delta m_{21}^2$ to date. 
Earlier, CHOOZ~\cite{chooz_last} and Palo Verde~\cite{pv_final} provided what are still the best 
upper limits on the mixing angle $\theta_{13}$.   A variety of new experiments are being constructed 
to perform more sensitive measurements of the mixing angle 
$\theta_{13}$~\cite{anderson,dbl_chooz,daya_bay,angra,reno}.
Anti-neutrino detectors at very short 
distance are being explored with the purpose of detecting coherent neutrino-nucleon 
scattering (C$\nu$NS)~\cite{CnuNS}, and for testing plutonium diversion at commercial 
reactors~\cite{songs}. A proper understanding of the systematic effects involved in the 
modeling of the reactor anti-neutrino flux and energy spectrum is essential for 
these experiments. Moreover, as most experiments measure $\bar{\nu}_{e}$s 
from more than one reactor, it is important to understand not only the magnitude of these effects, 
but also the correlations between uncertainties from different sources.

Except for the case of C$\nu$NS, which will not be addressed further here, 
reactor anti-neutrinos are usually detected using the inverse-$\beta$
reaction $\bar{\nu}_e + p \rightarrow n + e^+$ whose correlated signature helps reduce backgrounds
but is only sensitive to $\bar{\nu}_e$s with energy above 1.8 MeV. 
The kinetic energy of the positron gives a measure of the incoming $\bar{\nu}_e$ energy.
Neutrino oscillation parameters 
may be extracted from the data by fitting the observed $\bar{\nu}_{e}$ spectrum $\frac{dn_\nu}{dE_\nu}$ 
to the following equation:
\begin{equation}
\frac{dn_\nu}{dE_\nu} = \sum_k^{\textrm{reactors}} N_p \epsilon(E_\nu) \sigma(E_\nu) \frac{P_{ee}(E_\nu, L_k)}{4 \pi L_k^2} S_k(E_\nu).
\label{eq:NuebarSpecCalc}
\end{equation}
Here, $N_p$ is the number of target protons, 
$\epsilon(E_\nu)$ is the energy-dependent detection efficiency, 
$\sigma(E_\nu)$ is the detection cross-section, 
$P_{ee}(E_\nu, L_k)$ is the oscillation 
survival probability for $\bar{\nu}_e$s traveling a distance $L_k$ from reactor $k$ to the detector, 
and $S_k(E_\nu)$ is the $\bar{\nu}_e$ spectrum emitted by reactor $k$.
The time variation of the flux may be
included in the fit to discriminate the time-varying reactor signal from
constant backgrounds. 
Equation~\ref{eq:NuebarSpecCalc} should technically be multiplied by the
detector resolution function and integrated over $E_{\nu}$, but
this detail is inconsequential for our analysis.

Nuclear power reactors operate on the principle that the fission of U and Pu isotopes and the 
subsequent decays of their daughter fragments release energy, generating heat.  
Large Q-value $\beta$-decays of unstable fission fragments are primarily 
responsible for the $\bar{\nu}_e$ emission of nuclear reactors. Decays of long-lived 
isotopes in the nuclear fuel and in spent fuel elements stored at the reactor site contribute 
at the sub-percent level to the detected $\bar{\nu}_e$ flux; this contribution has
been treated elsewhere~\cite{kopeikin2001} and is beyond the scope of this paper. Considering only fission 
reactions, the final term in Equation~\ref{eq:NuebarSpecCalc} may be expanded as
\begin{equation}
S(E_\nu) = \sum_i^{\textrm{isotopes}}  f_i \left( \frac{dN_{\nu i}}{dE_\nu} \right),
\label{eq:ReactorEmissionSpec}
\end{equation}
where $dN_{\nu i} / dE_{\nu}$ is the $\bar{\nu}_e$ emission spectrum per fission of isotope $i$, 
and $f_i$ is the number of fissions of isotope $i$ during the data 
taking period. Since $>$99.9\% of the energy- and $\bar{\nu}_e$-producing fissions are from $^{235}$U, 
$^{238}$U, $^{239}$Pu, and $^{241}$Pu~\cite{lester_thesis} (see Figure~\ref{fig:pv_fission_rates}), 
the summation in Equation~\ref{eq:ReactorEmissionSpec} is
performed over just these 4 isotopes.  For $^{235}$U, $^{239}$Pu, and $^{241}$Pu, 
the emitted $\bar{\nu}_e$ spectra $dN_{\nu_i} / dE_\nu$ 
are derived from $\beta$-spectrum measurements of the fissioning of the isotopes 
by thermal neutrons~\cite{schreck1985,hahn1989}. 
The uncertainty in the $\bar{\nu}_e$ spectral shape derived from these
measurements was investigated recently in~\cite{vogel2007}.
For $^{238}$U, no measurements are available, 
so theoretical calculations of its $\bar{\nu}_e$ emission must be used~\cite{vogel1981}. 

The $f_i$ can be obtained from detailed simulations of the reactor core throughout the 
data taking period. 
The output $f_i$ must obey the thermal energy constraint
\begin{equation}
\label{eq:PowerSum}
W_{th} = \sum_{i=1}^{\textrm{isotopes}} f_i e_i,
\end{equation}
where $W_{th}$ is the total thermal energy produced by the reactor during
the time period considered, and $e_i$ denotes the energy released per fission of isotope $i$.
To reduce sensitivity to errors in the simulation codes, the codes are typically used to obtain 
not the $f_i$ directly but the fission fractions $f_i / F$, where $F = \sum_i f_i$. 
Measurements of the total generated thermal power taken regularly 
during reactor operation may then be used to 
obtain the total numbers of fissions of each isotope using Equation~\ref{eq:PowerSum}.

The appropriate simulation codes are cumbersome to run, and are often proprietary, 
and therefore are not always available for direct use by scientific
collaborations. 
Obtaining the $f_i/F$ therefore requires special agreements with reactor operators.
Depending on the terms of the
agreement, fine grained time-dependent fission fractions may be obtained. 
Fine-grained data for $W_{th}$ are commonly available, since such information must be 
made available to nuclear regulatory bodies. Reactor $\bar{\nu}_e$ experiments 
then use this information to calculate the $\bar{\nu}_e$ signal using an equation of the 
following form:
\begin{equation}
\label{eq:NuebarSpecCalcMod}
n_\nu = \frac{W_{th}}{{\sum_i (f_i/F) e_i}} \sum_{i=1}^{\textrm{isotopes}} \left(\frac{f_i}{F}\right)n_i,
\end{equation}
where for simplicity we have assumed a single reactor, and have integrated
over $E_{\nu}$. The terms in Equations~\ref{eq:NuebarSpecCalc} 
and~\ref{eq:ReactorEmissionSpec} not appearing in Equation~\ref{eq:NuebarSpecCalcMod} 
have been grouped into the $n_i$, the number of detected $\bar{\nu}_e$ per fission of isotope $i$.

Most recent reactor $\bar{\nu}_e$ experiments, including
KamLAND~\cite{kamland_prl,kamland_prl_2,kamland_prl_3},
CHOOZ~\cite{chooz_last}, Palo Verde~\cite{pv_final} and Bugey~\cite{bugey1994}, 
have used Equation~\ref{eq:NuebarSpecCalcMod}, or variations on it, to perform their
signal calculations.
In the first two experiments, the systematic uncertainty in the signal calculation was taken to be
the quadratic sum of the uncertainty in the reactor power measurements which yield $W_{th}$
and some estimate of the error due to the uncertainty in the fission fractions $f_i/F$.  In the case
of Palo Verde this latter component was estimated from comparisons between
calculated and measured 
isotopic concentrations in spent fuel elements~\cite{lester_thesis}.
In the case of KamLAND this component was taken from comparisons of 
a simplified reactor model to detailed 
simulations~\cite{kamland_prl,kamland_prl_2,kamland_prl_3,KLReactorSignal},
which assumes that the uncertainty of the simplified reactor
model significantly exceeds that of the detailed simulations.
However, no systematic uncertainty was assigned to the detailed reactor
simulations.
The CHOOZ experiment~\cite{chooz_last} leveraged the short baseline Bugey
observation to limit the reaction cross section uncertainty, 
scaling their result to agree with the Bugey-measured cross section per
fission. This treatment allowed CHOOZ to significantly reduce the error
associated with the $\bar{\nu}_e$ emission spectra ($dN_{\nu}/dE_{\nu}$).
However, their method is legitimate only to the extent that the reactor
simulations employed by the CHOOZ and Bugey reactor operators reliably model the
different reactors with their specific fuel compositions and operation histories.
In all cases, the combined systematic error was estimated to be within
2-3\%. At this level, the different methods employed by each experiment are
likely not overly aggressive.

With detector-based uncertainties in current and next-generation reactor $\bar{\nu}_e$ experiments
approaching the $\sim$1\% scale, a
more careful treatment of these reactor-specific uncertainties is necessary, 
particularly for experiments with multiple reactor sources in which 
correlations become important. 
In this paper we outline such a detailed treatment of these uncertainties  and 
demonstrate its application to a counting analysis of $\bar{\nu}_e$'s from single and 
multiple commercial reactor sources.
In Section~\ref{power_ch} we discuss in detail the uncertainty in 
reactor thermal power measurements.
In Section~\ref{compo_ch} we examine 
the Monte-Carlo estimation of the $f_i$ and, expanding on the treatment
in~\cite{lester_thesis} and~\cite{zelimir_thesis}, derive uncertainties for these based on a
large body of spent fuel isotopic concentration comparisons used to verify the codes.
In Section~\ref{analysis_ch} we combine the uncertainties from $W_{th}$ and the $f_i$ 
to estimate their contribution to the uncertainty in the
anti-neutrino yield. We also discuss the applications of our calculation to specific 
experimental configurations, including multiple reactor sources.
We draw our conclusions in Section~\ref{conclusion}. 

Our analysis addresses issues common 
to large commercial power reactors, especially Pressurized-Water-Reactors (PWRs) 
and Boiling-Water-Reactors (BWRs), using low-enrichment fuel.
Explanation of the operation of PWRs and BWRs may be found in~\cite{reactor_book}.
Modern neutrino experiments are almost exclusively using $\bar{\nu}_e$ from such reactors.  
Minor contributions from other reactor types are not considered here.
All uncertainties are given at the 68.3\% confidence level.

\section{Thermal Power Uncertainty}\label{power_ch}

The most accurate measurement of a reactor's thermal power is given by a calculation 
of the energy balance around the reactor
vessel (BWR) or steam generator (PWR). This requires accurate measurements of feed-water mass flow and
temperature, steam enthalpy and moisture content, and reactor coolant-cycle heat gains and
losses~\cite{crossflow}. For the PWRs at Diablo Canyon Power Plant (DCPP) 
(California, USA), the thermal balance is written as
\begin{equation}
Q_{C} = Q_{S} + Q_{LTND} + Q_{R\&C} - Q_{RCP} - Q_{PZ}.\label{eq:eq_4}
\end{equation} 
$Q_{C}$ is the core thermal output, the time-integral of which gives
$W_{th}$. The largest component of it, $Q_{S}$,
is the power extracted from the steam produced directly in the steam generator.
Smaller corrections are represented by: 
$Q_{LTND}$, the power lost in the water clean-up system;
$Q_{R\&C}$, the power losses to the external environment due to radiation and convection;
$Q_{RCP}$, the contribution to $Q_S$ from heating of the working fluid by the circulation pumps;
and $Q_{PZ}$, the contribution to $Q_S$ from heating of the working fluid by the pressurizers.
The correction terms together account for 0.3-0.4\% of the total $Q_{C}$.

A similar thermal balance may be written for other PWRs as well as for BWRs. 
In the case of BWRs, the main component is also in the form of steam produced
by the reactor. For both reactor types, $Q_S$ is evaluated according to
\begin{equation}
\label{eq:eq_3}
Q_{S} = m_s\;(h_{out} - h_{in}),
\end{equation}
where $m_s$ is the mass flow rate of the feed-water to the steam generator (PWR)
or reactor vessel (BWR), and $h_{out} - h_{in}$ is the specific enthalpy rise in the steam 
generation. We will discuss the uncertainties in these terms below.
The uncertainties on the terms other than $Q_{S}$ in the thermal balance equation (Equation~\ref{eq:eq_4})
contribute negligibly to the uncertainty of the core thermal power and will not 
be addressed further here.

The enthalpy rise is calculated, using steam tables~\cite{steam_tables},
from inlet and outlet values of the pressure and temperature, and from the moisture content
on the secondary side of the steam generator or reactor vessel.
The enthalpy uncertainty has several contributions deriving from measured
and calculated quantities. The errors in the temperature and pressure measurements 
are assumed to be random after correction for known systematic contributions.
The moisture content of the steam is generally known to grow during the
operation cycle because of reduced moisture removal efficiency of the
steam separators due to a slow deposition of eroded metal particles. 
The moisture content uncertainty of the steam may be treated as a systematic uncertainty 
since all reactors undergo this aging process.  The calculation of the
enthalpy rise from these inputs using steam tables contributes an additional systematic uncertainty of 
$<$0.2\%~\cite{steam_tables_uncertainty}.

As an example, Table~\ref{tab:table3} shows the uncertainties of the quantities relevant
to the enthalpy calculation for DCPP.  The moisture content at DCPP's steam 
generators grows with time, as mentioned above; the number in Table~\ref{tab:table3} is an 
average over the period of operation. Over longer periods the
moisture content may grow, as explained above, by a factor of two over the level
represented in Table~\ref{tab:table3}, but even then it will
comprise a sub-dominant contribution to the error budget.
A similar analysis for Beaver Valley Unit 2 yielded enthalpy-related
contributions to the thermal power uncertainty totaling
0.16\%~\cite{BeaverValley}.  A more general discussion
in~\cite{takamoto2007} uses an enthalpy uncertainty of 0.24\%.  
For a generic reactor or when a more detailed analysis is unavailable,
we suggest conservatively assuming 0.15\% random uncertainty 
(for the $p$ and $T$ measurements) and 0.2\% systematic uncertainty 
(due to the moisture content and use of steam tables) for the enthalpy rise.

\begin{table}
\caption{\label{tab:table3} Uncertainty contribution to the enthalpy from measured 
input quantities at DCPP. 
The change of the enthalpy vs given quantity, ($\Delta$Enthalpy/$\Delta$Quantity),
is calculated using steam tables~\cite{steam_tables}.
$p_{in}$, $T_{in}$ and $p_{out}$ are the inlet pressure, the inlet temperature, and
the outlet pressure, respectively. The present moisture carryover value at DCPP is
assumed in the table. The quadratic sum conservatively assumes a full 0.2\% error
related to the use of steam tables to calculate the enthalpy rise.
}
\begin{ruledtabular}
\begin{tabular}{lcccc}
\multirow{2}{*}{Quantity}  & Typical  & Quantity  & $\underline{\Delta\mathrm{Enthalpy}}$     & Enthalpy           \\ 
                 & Value  & Error [\%]         & $\Delta$Quantity                          & Error [\%]          \\\hline
$p_{in}$     &  6.9 MPa   & 0.50               & 0.002                                     & 0.001              \\
$T_{in}$     &   221$^\circ$C & 0.12               & 1.153                                     & 0.138              \\
$p_{out}$   &   5.6 MPa   & 0.94               & 0.018                                     & 0.017              \\
Moisture content  & 0.99 & 0.05               & 0.562                                     & 0.028              \\
Steam tables      &     &               &                                           & $<$0.2              \\
\multicolumn{3}{l}{{Quadratic Sum}}           &                                       & 0.25              \\
\end{tabular}
\end{ruledtabular}
\end{table}

The uncertainty in $m_s$ gives the largest contribution to the error 
in the thermal power~\cite{ray_f_0,lynnworth}.
In order to provide as accurate a measurement as possible,
the flow rate is measured in a long, straight section of pipe where fully
developed turbulent flow is established.  Different types of flow meters may be employed there. 
Traditionally the system is instrumented with nozzle-based flow
meters operating on the principle of the Venturi effect.  An early analysis 
yielded a total flow uncertainty for these meters of 1.7\%, of which 1.6\% is
attributable to random errors, while the systematic uncertainty is determined to be 0.5\%~\cite{GETAB}.
According to more recent experience at DCPP, properly calibrated and maintained
Venturi meters can be operated with an initially low total uncertainty of about 0.7\%. 
However, the accuracy of the Venturi meters fluctuates with time due to
material deposition, or ``fouling''~\cite{ray_f_0}, 
which leads to smaller cross sectional area for the flow through the device and 
results in indicated flow rates exceeding the true value.
Depending upon the nozzle style and the pH factor of the feed-water, this fouling
can grow as high as 3\% over periods of a few months to a few years. 
In-situ comparisons between Venturi meters and more modern
instrumentation (discussed below) shows an average systematic difference of +0.6\%, and a statistical
spread of $\sim$1.4\%~\cite{nozzle_bias, nozzle_bias_2}.

Fouling in Venturi flow meters results in economic losses for the
nuclear power industry.  Nuclear reactor regulations in the US
and elsewhere require operators to apply a safety margin accounting for the
uncertainty in the thermal power measurement when performing
loss-of-coolant accident and emergency core cooling system performance
analyses. However, the regulatory practice varies from country to country~\cite{IAEA-impl}.
The size of the safety margin is 2\% in the US, but may
be reduced if the thermal power measurement uncertainties can be
demonstrated to be smaller than that level~\cite{us_reg}.  For
reactors instrumented with Venturi meters, for which lower uncertainties
cannot be demonstrated as a result of fouling, this in effect leads
to operation at 2\% below the plant's licensed operating power
limit.  A similar 2\% safety margin is required in Japan~\cite{jp_reg}.
Moreover, fouling itself results in an overestimation of the thermal
power, leading to actual operation up to a few percent below the
indicated level, and thus a lower electric power output. This two-fold
impact of fouling gives the nuclear power industry a high incentive to
deploy a replacement technology with smaller, well-understood
uncertainties.

Using a new generation of ultrasonic flow meters (UFM) not affected by 
fouling, the thermal power uncertainty may be significantly reduced.
UFMs use an electronic transducer with no differential pressure elements.
There are currently two types of UFM systems routinely used in the nuclear electric generation 
industry to accurately calculate feed-water flow in high purity water~\cite{ray_f}.
One style is the transit-time meter, while the other is the cross-flow meter.
The transit-time UFM emits ultrasound signals diagonally through the fluid
upstream and downstream along the same path. The measured time difference
between the two paths is proportional to the velocity of the fluid in the pipe.
Accuracies of 0.2-0.5\% are reported~\cite{caldon, estrada} for transit-time type UFMs. 
The cross-flow UFM measures the time taken by a unique pattern
of turbulent eddies in the fluid to travel between
two pairs of transducers at some known distance apart along the pipe.
The cross-flow devices are mounted externally to the pipe.
The cross-flow UFM with its associated hardware and software is able to 
achieve a flow measurement uncertainty of 0.25\% or better, although
the ability to achieve these uncertainties in real reactor conditions has
been questioned~\cite{us_reg2}.
This type of meter is employed at DCPP, with an uncertainty evaluated
in-situ within the 0.4-0.7\% range.
The principles of operation of both transit-time and cross-flow UFMs 
are described in detail in \cite{flow_book}.

The measurement uncertainties of transit-time UFMs and their breakdown into statistical
and systematic components were studied in detail by
Estrada~\cite{estrada} of Caldon Inc., a flow meter manufacturer
for the nuclear industry.
This work found total uncertainties in the measured mass flow rates of 0.45\% for
models with externally mounted transducers, and 0.20\% for intrusive-type
models with 4 pairs of transducers. We refer to these as Type I and Type II UFMs, respectively.
These uncertainties originate from the differences between the water axial velocity
profile in the test facility and in the measurement system in a plant, the uncertainty
in the measurement of the ultrasound acoustic path in the water pipe, 
the imperfect knowledge of the dimensions of the measurement systems, and the
uncertainty in the
measurement of the time of flight of the acoustic pulses, including non-fluid delays.
The water flow uncertainties are divided into random and systematic contributions
in Table~\ref{tab:table4}. 

We note that the 0.45\%
uncertainty based on Estrada's study is consistent with the 0.4-0.7\% uncertainty
estimated by DCPP for cross-flow meters.
Several other assessments of mass flow rate uncertainties,
also consistent with these values, 
may be found in the literature. The generic iscussion in \cite{takamoto2007}, for example, 
quotes feed-water flow uncertainties of 0.4\%. For a generic UFM, we
will use the values listed for the Type I transit-time UFM, which has the largest
estimated systematic uncertainty, and gives a total error
consistent with values considered by the US Nuclear Regulatory Commission
in a discussion on the use of UFMs~\cite{us_reg2}.

\begin{table}
\caption{\label{tab:table4} Typical flow meter uncertainties.
Type I corresponds to transit-time (TT) UFMs with ultrasonic transducers
externally mounted to the feed-water pipe.
Type II corresponds to transit-time UFMs with ultrasonic transducers 
in-line with the pipe via a spool piece that integrates four pairs of
ultrasonic transducers, forming four chordal paths.
Type I-II UFM water flow uncertainties are based on Caldon Inc. 
experience~\cite{estrada}. 
The flow meter employed at DCPP is of the cross-flow (CF) type. }
\begin{ruledtabular}
\begin{tabular}{lccc}
Flow Meter    & Random [\%] & Syst. [\%] & Total [\%] \\ 
\hline
Venturi       & 1.4        & 0.6        &  1.5   \\
Type I  TT UFM    & 0.2         & 0.4        &  0.45      \\
Type II TT UFM   & 0.09        & 0.18       &  0.20      \\
DCPP CF UFM      & 0.3-0.6     &   0.1-0.2         &  0.4-0.7   \\
\end{tabular}
\end{ruledtabular}
\end{table}

The errors on the enthalpy rise and $m_s$ may be added in quadrature to
obtain the full error on $Q_S$.
Combining the mass flow uncertainty (as given in Table~\ref{tab:table4} for Type I transit-time UFMs) with an enthalpy 
uncertainty of 0.25\%, we get a total thermal power uncertainty of 0.51\%, 
of which 0.45\% is systematic and 0.25\% will vary from reactor-to-reactor.
Typical values for the quantities describing the heat balance and 
their error are given in Table~\ref{tab:table1} for DCPP, using two PRWs
equipped with cross-flow type UFMs.  
An internal DCPP study characterized about a quarter of the thermal power uncertainty as correlated.
The error in the
determination of $Q_S$, which is the only non-negligible component of $Q_C$ and
hence determines the uncertainty on $W_{th}$, is dominated by the uncertainty in $m_s$. 
For precise evaluation of the errors, the enthalpy uncertainty has to be taken into account.

\begin{table}
\caption{\label{tab:table1} Typical thermal balance quantities
for the Diablo Canyon PWR reactors. }
\begin{ruledtabular}
\begin{tabular}{llc}
Quantity                      & Typical Value          & Error [\%]  \\ \hline
$m_s$                         & 1887 kg/s              &0.4-0.7\\ 
$\Delta h = h_{out} - h_{in}$ & $1.819\cdot 10^{6}$ J/kg & 0.25   \\ \hline
$Q_{S}$                      & 3433 MWt               & $\lesssim$0.7\\
$Q_{LTND}$                    & 1.83 MWt               &  nil    \\
$Q_{R\&C}$                    & 0.65 MWt               &  nil    \\
$Q_{RCP}$                     & 14 MWt                 &  nil    \\
$Q_{PZ}$                      & 0.21 MWt               &  nil    \\
$Q_{C}$                       & 3421 MWt               & $\lesssim$0.7\\
\end{tabular}
\end{ruledtabular}
\end{table}

It is worth mentioning that most feed-water flow applications require 
two or more flow measurements. In PWRs, for example, the flow is measured
in each steam generator.
In BWRs, the flow is normally measured in each of two main feed headers. In
these cases, the random uncertainties 
may be reduced by a factor of $1/\sqrt{\textrm{number of generators}}$
or $1/\sqrt{2}$, respectively. 

It is also worth noting that estimates of flow rate measurement uncertainties
to date have been performed in laboratory settings at Reynolds
numbers of up to 10$^6$. However, Reynolds numbers in actual power
plants are as high as 10$^7$~\cite{takamoto2007, japan_up, epri}. The
uncertainty estimates described above are therefore extrapolations
from the lower Reynolds number evaluations.
Although the error associated with these extrapolations is expected to be small, 
new test facilities have been proposed to generate realistic flow conditions 
with a higher Reynolds number to test these extrapolations~\cite{furuichi2}.

\section{Uncertainties in Fission Calculations}\label{compo_ch}

During the power cycle of a nuclear reactor, the composition of the fuel changes as Pu isotopes 
are bred and U is depleted. At the end of the power cycle, some fraction of the fuel is 
replaced, and the remaining fuel elements are shuffled to optimize burnup. Knowledge of 
the detailed, time-dependent radionuclide content of the spent fuel throughout the power cycle 
is of interest to reactor 
operators and regulatory groups because it impacts shielding requirements, dose rate 
analysis, toxicity, waste storage and handling considerations, and criticality safety~\cite{ornl_2003}. 
To obtain this knowledge, reactor operators employ Monte-Carlo simulations of the 
reactor core. The simulations begin with an initial fuel composition and a physical 
model of the core assembly. An initial neutron flux solution for the core is computed based on a number 
of inputs, including the measured thermal power generation, and other 
operating parameters such as the pressure, temperature, and flow rate of the cooling system,
neutron moderation parameters, etc. Cross-section and decay data libraries are used 
to compute the rate of various interactions and update the fuel composition
as a function of position. 
This process is  iterated in a series of time-steps throughout the reactor power cycle. 

\begin{figure}
\begin{center}
\includegraphics[width=8.6cm]{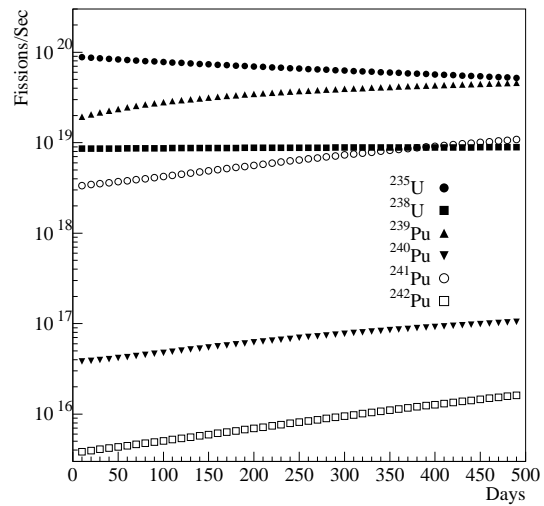}
\end{center}
\caption{\label{fig:pv_fission_rates} Fission rates for various isotopes during a typical 500 day fuel cycle for one of the Palo 
Verde Nuclear Generating Station reactors in Arizona (USA).  $>$99.9\% of the $\bar{\nu}_e$s 
are emitted by $^{235}$U,  $^{238}$U, $^{239}$Pu, and $^{241}$Pu~\cite{lester_thesis}.}
\end{figure}

As a by-product, these reactor core simulations may be made to output
the number of fissions $f_i$ needed for the estimation of the signal in reactor anti-neutrino experiments.
A plot of the time-dependent fission rates calculated for a typical power cycle of one of the reactors 
at the Palo Verde Nuclear Generating Station (PVNGS) in Arizona, USA, is shown in 
Figure~\ref{fig:pv_fission_rates}. Integrating over the power cycle gives the $f_i$
required by Equation~\ref{eq:NuebarSpecCalcMod} to compute the
$\bar{\nu}_e$ signal. 

\begin{figure}
\begin{center}
\leavevmode
\includegraphics[width=8.6cm]{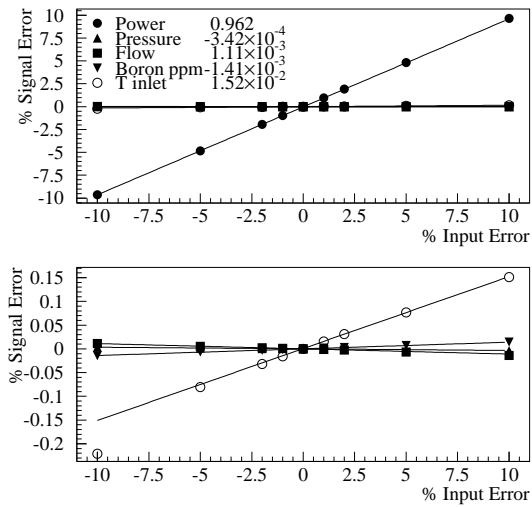}
\end{center}
\caption{\label{fig:pv_signal_error} The change in the expected $\bar{\nu}_{e}$-signal as a function 
of five inputs to the core simulation, taken from \cite{lester_thesis}. The numbers in the key are 
slopes of the fitted lines. The lower panel zooms in on the non-power inputs so that their variation
can be observed. }
\end{figure}

Uncertainties in these simulation codes originate from a variety of sources~\cite{gauld_2003}.
These sources include the uncertainties in the input parameters, uncertainties 
in the nuclear cross-section and decay data used by the codes, approximations made in the
modeling of the reactor core, and numerical approximations used in the computations 
themselves. 
Reference~\cite{lester_thesis} explored the input parameter uncertainty
contribution with the reactor core simulation code used at PVNGS~\cite{rocs,annals} 
by calculating the $\bar{\nu}_e$ signal variation for a given deviation of several of the
input parameters.
For the PVNGS simulation, the normalization to the measured thermal power
represented by Equation~\ref{eq:NuebarSpecCalcMod} was done within the
code.  The code was run for a set 
of 500-day power cycles with one parameter varied at a time. The results of the study 
are shown in Figure~\ref{fig:pv_signal_error}. The thermal power measurement has the largest effect, 
with a slope of 0.96.

The most common means reactor operators use to globally assess the performance of
the codes is to compare measurements of the isotopic content of spent
fuel elements with the values predicted by the simulations.
Such verifications have been performed for a number of codes
at a variety of PWRs and BWRs 
in the US, Japan and Europe~\cite{suyama_2002_1,suyama_2002_2,nakahara_2002,pnl}.
Samples are obtained from representative locations throughout the core, commonly
selected from regions that are difficult to simulate accurately, such as boundaries and hot-spots.
The assay technique employed is typically isotopic dilution mass
spectroscopy, for which errors of 1-3\% are typical for the
isotopes of interest~\cite{burk_thesis}; in more recent measurements, 
errors below 0.3\% have been acheived~\cite{ornl_2003}.

\begin{figure}
\begin{center}
\leavevmode
\includegraphics[width=8.6cm]{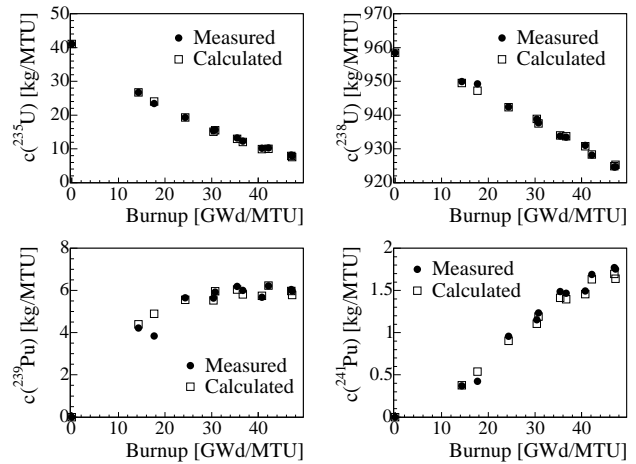}
\end{center}
\caption{\label{fig:takahama3_burnup_info} Measured vs. calculated concentrations of $^{235}$U,
$^{238}$U, $^{239}$Pu and $^{241}$Pu as a function of burnup for
fuel elements taken from
the Takahama Unit 3 PWR~\cite{ornl_2003}. See the text for a definition of the units.}
\end{figure}

Figure~\ref{fig:takahama3_burnup_info} shows an example comparison between
measured and calculated concentrations, in kilogram per metric tonne uranium (kg/MTU),
of $^{235}$U, $^{238}$U,
$^{239}$Pu and $^{241}$Pu as a function of burnup for a set of fuel elements
processed at a typical PWR~\cite{ornl_2003}. 
The quantity ``burnup'' in the abscissa is defined as the amount of energy (in GigaWatt days)
extracted from a fuel element per unit initial mass of uranium (in MTU).
The burnup is measured to within $\sim$3\% for each sample using the $^{148}$Nd 
method~\cite{nd_method}, in which the integral number of 
fissions, determined from the sum of the $^{148}$Nd fission product yields
for the four isotopes of interest, is multiplied by the average energy released per fission,
weighted according to the calculated fission fractions.
During a typical reactor cycle, fuel elements near the center
of the reactor core receive a higher burnup than those at the edges, due
to the higher neutron flux at the center. 
This spatial variation, along with the number of cycles over which the samples 
were processed, is responsible for the different burnup values 
achieved in the samples plotted in Figure~\ref{fig:takahama3_burnup_info}. 
During normal operations, fuel elements are processed over several power
cycles, and are shuffled between each power cycle, until optimal burnup is reached 
for all fuel elements. In practice, burnups up to $\sim$50 GWd/MTU are achieved in the 
typical operation of modern commercial reactors~\cite{doeburnup}.

\begin{figure}
\begin{center}
\leavevmode
\includegraphics[width=8.6cm]{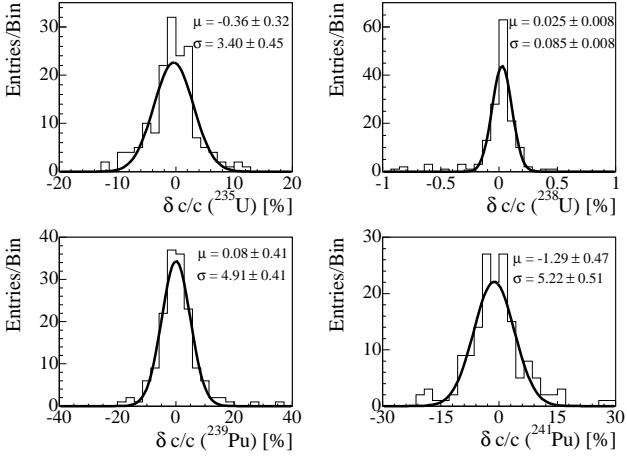}
\end{center}
\caption{\label{fig:1D_histo_fuel_deltaCoverC} The values $\delta c/c$ for  $^{235}$U, $^{238}$U, $^{239}$Pu 
and $^{241}$Pu, calculated for each of the 159 comparisons of fuel element samples taken from a number of PWRs and BWRs modeled by a variety of core simulation codes~\cite{ornl_1995, ornl_2003, ornl_1998, ornl_check_year, mox, trans}. }
\end{figure}

\begin{figure}
\begin{center}
\leavevmode
\includegraphics[width=8.6cm]{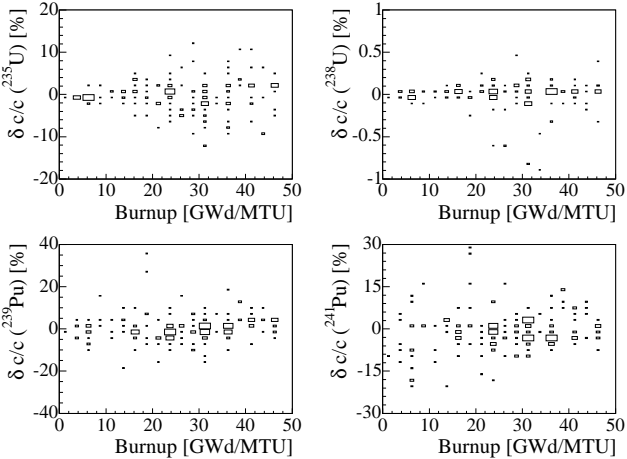}
\end{center}
\caption{\label{fig:2D_deltaCoverC_vs_burnup} The values $\delta c/c$ for  $^{235}$U, $^{238}$U, $^{239}$Pu 
and $^{241}$Pu as a function of burnup, calculated for each of the 159 fuel sample comparisons.}
\end{figure}

Fractional differences in the heavy isotope concentrations $\delta c_i/c_i$
were calculated from
the measured ($M$) and calculated ($C$) concentrations 
using the following sign convention:
\begin{equation}
\label{eq:signConvention}
\frac{\delta c}{c} = \frac{c_C - c_M}{c_M}.
\end{equation}
In Figure~\ref{fig:1D_histo_fuel_deltaCoverC} we plot the values of $\delta c_i/c_i$
($i$ = $^{235}$U, $^{238}$U, $^{239}$Pu, $^{241}$Pu), using 159 comparisons of 
fuel element samples taken from ten PWRs and BWRs, modeled by a variety of core simulation codes.  
Details of the codes and reactors included in our analysis are
listed with their references in Table~\ref{tab:dataset}.
Gaussian fits to the distributions are drawn for reference.
The individual distributions for PWRs and BWRs (not shown) are equivalent within
the available statistics.
The concentration comparisons also do not exhibit a strong trend with burnup,
as shown in Figure~\ref{fig:2D_deltaCoverC_vs_burnup}.

\begin{table}
\caption{\label{tab:dataset} Codes and reactors included in our analysis. In cases 
where multiple tests were performed on the same code/reactor using 
older cross section files or neglecting some reactor configuration or operation details, 
we used the comparisons for the most up-to-date data libraries and most detailed 
reactor model. 
}
\scriptsize
\begin{ruledtabular}
\begin{tabular}[t]{cccc} 
Ref. &  Reactor (Type) & $\mathrm{W_{th}}$ [MWt] & Code  \\ \hline
\\[-1ex]
 & & & SAS2H \\[-1ex] 
\raisebox{1.5ex}{\cite{ornl_2003}} & \raisebox{1.5ex}{Takahama Unit 3 (PWR)} & \raisebox{1.5ex}{2652} & HELIOS \\[1ex]
\hline
\\[-1ex]
& Calvert Cliffs Unit 1(PWR) & 2560 & \\[-1ex]
\raisebox{1.5ex}{\cite{ornl_1995}} & H.B. Robinson Unit 2 (PWR) & 2192 & \raisebox{1.5ex}{SCALE/SAS2H} \\[1ex]
\hline
\\[-1ex]
& Cooper (BWR) & 2381 & \\
\cite{ornl_1998} & Gundremmingen (BWR) & 801 & SAS2H \\
& JDPR (BWR) & 45 &\\[1ex]
\hline
\\[-1ex]
& Trino Vercellese (PWR) & 825 & \\[-1ex]
\raisebox{1.5ex}{\cite{ornl_check_year}} & Turkey Point Unit 3 (PWR) & 2300 & \raisebox{1.5ex}{SAS2H} \\[1ex]
\hline
\\[-1ex]
\cite{mox} & San Onofre (PWR) & 1347 & SAS2H \\[1ex]
\hline
\\[-1ex]
& & & SWAT \\[-1ex]
& \raisebox{1.5ex}{Takahama Unit 3 (PWR)} & \raisebox{1.5ex}{2652} & ORIGEN2 \\[-1ex]
\raisebox{1.5ex}{\cite{trans}} & &  & SWAT \\[-1ex]
& \raisebox{1.5ex}{Fukushima Unit 2 (BWR)} & \raisebox{1.5ex}{3293}  & ORIGEN2 \\[1ex]
\hline
\\[-1ex]
\cite{ornl_1996} & Calvert Cliffs Unit 1(PWR) & 2560 & ORNL-Assm \\[1ex]
\end{tabular}
\end{ruledtabular}
\end{table}

Unfortunately, these isotopic concentration comparisons only give indirect information on 
the uncertainty in the number of fissions. However, we can estimate $\delta f_i/f_i$ from 
$\delta c_i/c_i$ by considering the nuclear processes responsible for the changes
in isotopic concentrations through the lifetime of a typical fuel element.
The U isotopes are the most straightforward to diagnose, since U
breeding processes are negligible compared to U depletion processes.
The decrease of the U concentrations with burnup results from a combination of fission 
and isotopic conversions due to other nuclear processes (e.g.~n capture). 
The error on the total loss of U is 
the fractional error in the change in concentration after burnup $B$.
We conservatively assign the error on the total loss of U
entirely to fission processes, so that the fractional error on the 
number of U fissions is at most
\begin{equation}
\frac{\delta f_{\rm U}}{f_{\rm U}} = \frac{\delta(c_{\rm U}(0) - c_{\rm U}(B))}{c_{\rm U}(0) - c_{\rm U}(B)},
\label{eq:dfiAssumption}
\end{equation}
where $c_{\rm U}(0)$ and $c_{\rm U}(B)$ are, respectively, the initial and final concentrations
of either U isotope after burnup $B$. The initial $^{235}$U and $^{238}$U concentrations
($c_{\rm U}(0)$) are determined by the enrichment level of the fuel elements
under consideration and are known very precisely. Hence the
uncertainty $\delta(c_{\rm U}(0) - c_{\rm U}(B))$ is dominated by the uncertainty
in $c_{\rm U}(B)$, allowing us to approximate
\begin{equation}
\frac{\delta f_{\rm U}}{f_{\rm U}} \approx - \left( \frac{1}{c_{\rm U}(0)/c_{\rm U}(B) - 1} \right) \frac{\delta c_{\rm U}(B)}{c_{\rm U}(B)},
\label{eq:dccToDff}
\end{equation}
where the last term is the $\delta c/c$ plotted in 
Figure~\ref{fig:1D_histo_fuel_deltaCoverC}. As can be seen in 
Figure~\ref{fig:takahama3_burnup_info}, which is typical for fuel
processed in commercial PWRs and BWRs, the factor in parenthesis
in Equation~\ref{eq:dccToDff} is significantly different from one for $^{238}$U
but not for $^{235}$U, except at low burnup.

For the Pu isotopes the relationship between the $\delta f_{\rm{Pu}}/f_{\rm{Pu}}$ and the 
$\delta c_{\rm{Pu}}/c_{\rm{Pu}}$ is complicated by the presence of significant breeding processes. 
However, the sources of uncertainty in $\delta f_{\rm{Pu}} / f_{\rm{Pu}}$ are the same as those
in $\delta c_{\rm{Pu}}/c_{\rm{Pu}}$, indicating a similar magnitude.
Moreover, since the Pu concentrations in fresh fuel used at typical commercial reactors 
starts at ``zero'' and the fission rate is at
all values of $B$ proportional to the concentration $c_{\rm{Pu}}(B)$, we simply assume that 
$\delta f_{\rm{Pu}}/f_{\rm{Pu}} \approx \delta c_{\rm{Pu}}/c_{\rm{Pu}}$ directly.

The four values of $\delta f_i/f_i$ ($i$ = $^{235}$U, $^{238}$U, $^{239}$Pu, $^{241}$Pu) 
were calculated for each of the 159 comparisons of fuel element samples from the reactors
and core simulation codes listed in Table~\ref{tab:dataset},
using Equation~\ref{eq:dccToDff} for the U isotopes and letting
$\delta f_{\rm{Pu}}/f_{\rm{Pu}} \approx \delta c_{\rm{Pu}}/c_{\rm{Pu}}$ for the Pu isotopes.
The means and standard deviations of the distributions for each isotope were extracted
from Gaussian fits. The results are listed in Table~\ref{tab:dff_values}.
Note that with our sign convention defined in
Equation~\ref{eq:signConvention}, a positive $\delta f_i/f_i$
corresponds to an over-estimated fission count by the reactor core
simulations. 

\begin{table}
\caption{\label{tab:dff_values} Means ($\mu$) and standard deviations
($\sigma$) of $\delta f_i/f_i$ for commercial reactors.}
\begin{ruledtabular}
\begin{tabular}[t]{ccc} 
Fuel       &  $\mu$ $[\%]$     & $\sigma$ $[\%]$ \\ 
\hline
$^{235}$U  &  0.09 $\pm$ 0.16  & 2.0             \\ 
$^{238}$U  &  -0.81 $\pm$ 0.38  & 4.8             \\ 
$^{239}$Pu &  0.74 $\pm$ 0.45  & 5.7             \\
$^{241}$Pu &  -0.32 $\pm$ 0.48  & 6.0             \\ 
\end{tabular}
\end{ruledtabular}
\end{table}

\begin{figure}
\begin{center}
\leavevmode
\includegraphics[width=8.6cm]{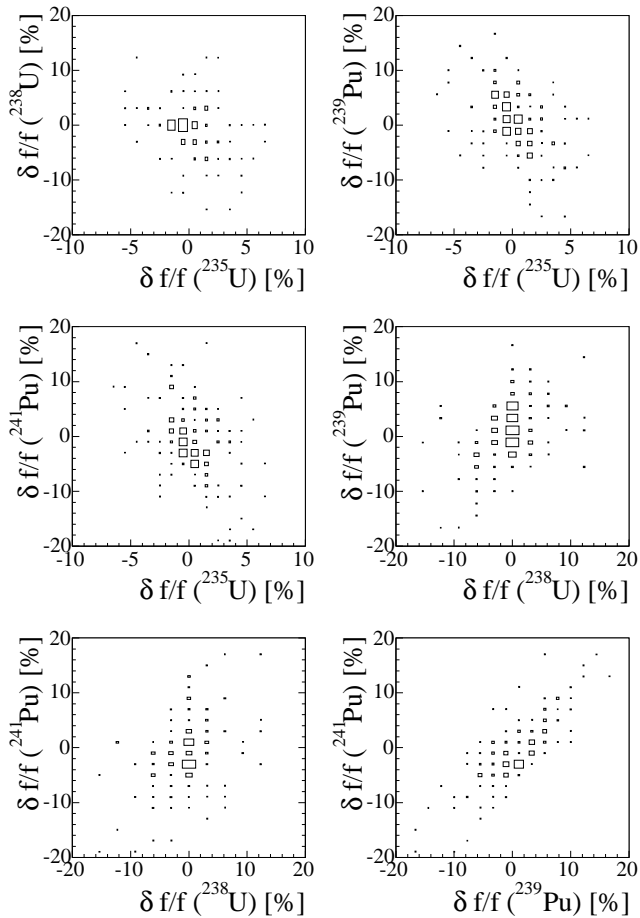}
\end{center}
\caption{\label{fig:fig_2D_histos} $\frac{\delta f_i}{f_i}$ dependencies for $^{235}$U, 
$^{238}$U, $^{239}$Pu, and $^{241}$Pu.}
\end{figure}

Correlations between the $\delta f_i/f_i$ were studied as well.  
Figure~\ref{fig:fig_2D_histos}
plots the $\delta f_i/f_i$ for pairs of fuel isotopes for all 159 fuel samples comparisons.
Correlation coefficients $\alpha_{ij}$ are obtained by fitting these distributions
to 2-dimensional Gaussians:
\begin{equation}
G(\Delta_i, \Delta_j) = A e^{-\frac{1}{2} \frac{1}{1-\alpha_{ij}^2} \left( \Delta_i^2 +\Delta_j^2 - 2\alpha_{ij}\Delta_i \Delta_j \right)},
\label{eq:TwoDGauss}
\end{equation}
where $\Delta_i \equiv (\delta f_i/f_i - \mu_i)/\sigma_i$ and $A$
is a normalization constant. In the fit, $\mu_i$ and $\sigma_i$ are
allowed to float, resulting in fit values consistent in all cases
with the values listed in Table~\ref{tab:dff_values}. The fit values
of $\alpha_{ij}$ are listed in Table~\ref{table:table_corr}.

\begin{table}
\caption{\label{table:table_corr} Correlation coefficients $\alpha_{ij}$
obtained from 2-dimensional Gaussian fits applied to the distributions
from Figure~\ref{fig:fig_2D_histos} }
\begin{ruledtabular}
\begin{tabular}{lcccc} 
           &  $^{235}$U & $^{238}$U & $^{239}$Pu & $^{241}$Pu \\ 
\hline
$^{235}$U   & 1             & -0.29     &  -0.62     &  -0.48     \\
$^{238}$U   & -0.29      & 1            &  0.38      &  0.39      \\
$^{239}$Pu & -0.62      & 0.38      &  1            &  0.84      \\
$^{241}$Pu & -0.48      & 0.39      &  0.84      &  1         \\ 
\end{tabular}
\end{ruledtabular}
\end{table}

We find a weak anti-correlation between $^{235}$U and all of the other isotopes.
This is expected since $^{235}$U dominates the thermal power production of
the reactor, and so to maintain the energy balance, the over-fissioning of
this isotope must be accompanied by an under-fissioning of the other isotopes
(and vice-versa). The fact that the largest anti-correlation is with $^{239}$Pu, the 
next largest energy producing heavy isotope after $^{235}$U, strengthens this argument.

We also find a strong correlation between the two Pu isotopes,
which can be explained by the fact that $^{239}$Pu is a precursor to $^{241}$Pu
in one of three main isotope transformation chains within the reactor
core~\cite{nakahara_1990}:
\begin{equation}
\label{eq:238U_chain}
^{238}{\rm U} \rightarrow ^{239}\hspace{-1.5mm}{\rm Pu} \rightarrow ^{240}\hspace{-1.5mm}{\rm Pu} \rightarrow ^{241}\hspace{-1.5mm}{\rm Pu} \rightarrow...\rightarrow ^{244}\hspace{-1.5mm}{\rm Cm}
\end{equation}
Since there is little or no Pu in a fuel element at zero burnup,
over-fissioning of a Pu isotope is associated with its overproduction
in the simulation via this isotope transformation chain. If one Pu
isotope is overproduced, it is likely that the other will also be
overproduced, resulting, in more fissions from both isotopes.

The correlation between $^{238}$U and Pu isotopes is a bit harder to diagnose.
However, $^{238}$U is only a minor player in the thermal power production,
comprising only $\sim$10\% of the output, so a strong anti-correlation is not 
expected as it is for $^{235}$U. 
And while $^{238}$U is the parent of the transformation chain represented
by Equation~\ref{eq:238U_chain}, that chain represents a series
of neutron captures, whereas the correlations listed in Table~\ref{table:table_corr}
correspond to fissions. The over-fissioning of $^{238}$U does not
a-priori imply a greater neutron capture rate to produce more of the Pu isotopes
(so that more are available to fission), and vice-versa. The true $^{238}$U-Pu correlation
is a combination of these competing effects.

\section{Uncertainty in the anti-neutrino signal calculation}\label{analysis_ch}

We now examine how to propagate the uncertainties in $W_{th}$ and the $f_i$,
found in the previous sections, into the anti-neutrino signal uncertainty.
This is not straightforward because of the significant correlation between
$\delta W_{th}$ and the $\delta f_i$, as
indicated by Equation~\ref{eq:PowerSum}.
To first order, an increase in the
thermal power by some factor implies an increase in the $f_i$ by the same
factor. In fact, if we use the $\delta f_i/f_i$ obtained from each
spent fuel element discussed in the previous section to
extract the thermal power uncertainty via standard error propagation on Equation~\ref{eq:PowerSum}, 
we obtain a distribution of $\delta W_{th}/W_{th}$ with width 2.5\%,
consistent with the expectation for a set of reactors instrumented
with Venturi-type flow meters of varying ages.

The point of the thermal power normalization in
Equation~\ref{eq:NuebarSpecCalcMod} is to break
this first-order correlation between the $f_i$ and $W_{th}$,
and work with the fission fractions $f_i/F$ rather than the bare $f_i$.
But even the $f_i/F$ have a residual correlation with the thermal
power that is due to the fact that U isotopes are depleted while Pu isotopes
are bred during reactor operation. An over-estimate of the power will lead to
an over-estimate of the Pu production, and hence a higher estimated fraction
of fissions from the Pu isotopes. This is significant for the $\bar{\nu}_e$ signal
estimation because, by coincidence, the $e_i$ are slightly higher and the $n_i$ 
are slightly lower for the Pu isotopes than for the U isotopes. Since the $e_i$
appear in the denominator of Equation~\ref{eq:NuebarSpecCalcMod} and the
$n_i$ appear in the numerator, the effect is to lower slightly the signal estimate.

Hence we see that the contributions to the signal uncertainty from the errors on 
$W_{th}$ and the $f_i/F$ are slightly anti-correlated. It is this anti-correlation that
is responsible for the slope of ``Power'' in Figure~\ref{fig:pv_signal_error}
being slightly less than one. However, the anti-correlation is only slight, and
moreover implies a weak cancellation between the uncertainty contributions
evaluated from the independent variation of $W_{th}$ and the $f_i/F$. 
To maintain simplicity in our analysis while remaining conservative in the error propagation, 
we ignore this anti-correlation
and treat the errors due to these terms as being uncorrelated. We thus
evaluate the uncertainty on the $\bar{\nu}_e$ signal, $\sigma_\nu$,
according to the simple quadratic sum of contributions from each of the terms 
appearing in Equation~\ref{eq:NuebarSpecCalcMod}:
\begin{equation}
\label{eq:NuebarRateUncertainty}
\sigma_{n_\nu}^2 = \sigma_W^2 + \sigma_f^2+ \sigma_e^2 + \sigma_{other}^2.
\end{equation}
Since Equation~\ref{eq:NuebarSpecCalcMod} is linear in $W_{th}$, the first term, $\sigma_W$, 
is given by the thermal power uncertainty, $\delta W_{th}/W_{th}$, discussed
in detail in Section~\ref{power_ch}. The contribution $\sigma_f$ from the fission calculation
uncertainties, $\delta f_i/f_i$ is more complex and will be described below. 
The third term, $\sigma_e$, is the contribution due to the uncertainty in the $e_i$.
The last term, $\sigma_{other}$, 
represents uncertainties in detector-specific components and 
other terms (such as the $dN_\nu/dE_\nu$) that have
been explored elsewhere. This term will not be addressed further here.

Before addressing $\sigma_f$, we briefly note that $\sigma_e$ can be evaluated 
from error propagation on Equation~\ref{eq:NuebarSpecCalcMod}. Assuming that 
the errors on each of the
$e_i$ are uncorrelated, we obtain
\begin{equation}
\label{eq:EiUncertainty}
\sigma_e^2 = \frac{\sum_i (f_i/F)^2 \delta e_i^2}{[\sum_i (f_i/F) e_i]^2}
\end{equation}
The values of the $e_i$ and their uncertainties $\delta e_i$ are given in 
\cite{zacek_1986,declais_1994} 
as $e(^{235}\rm{U}) = 201.7\pm 0.6$ MeV, $e(^{238}\rm{U}) = 205.0\pm 0.9$ MeV, 
$e(^{239}\rm{Pu}) = 210.0\pm 0.9$ MeV, and $e(^{241}\rm{Pu}) = 212.4\pm 1.0$ MeV.
To evaluate $\sigma_e$, we need to choose values for the fission fractions, $f_i/F$. 
As examples, Table~\ref{tab:reactor_info} shows the fission fractions for
the three reactor units of PVNGS (PV1, PV2, and PV3), as well as averaged fission
fractions of the reactors in Japan (KL), observed by the KamLAND experiment.
The fission fraction values vary from reactor to reactor depending
on the initial enrichment level and power history of the fuel.
Table~\ref{tab:reactor_info} gives the corresponding value of $\sigma_e$, 
calculated with Eq.~\ref{eq:EiUncertainty}. 
In all cases considered, the value of $\sigma_e$ is $\sim$0.2\%.

Correlations between the $\delta f_i/f_i$ make an analytical estimate of $\sigma_f$  difficult. 
Instead we leverage the measured $\delta f_i/f_i$ from Section~\ref{compo_ch} to
compute, for each of the 159 spent fuel element comparisons, a value of $\sigma_f$
according to
\begin{eqnarray}
\sigma_f & \equiv & \left(\frac{\delta n_\nu}{n_\nu}\right)_f  \nonumber \\
         & = & \frac{1 + \left(\sum_i \frac{f_i}{F} \frac{\delta f_i}{f_i} n_i\right) / \left(\sum_i \frac{f_i}{F} n_i\right)}{1 + \left(\sum_i \frac{f_i}{F} \frac{\delta f_i}{f_i} e_i\right) / \left(\sum_i \frac{f_i}{F} e_i\right)} - 1. \nonumber \\
\label{eq:dnn_f}
\end{eqnarray}
This equation may be obtained by replacing the
$f_i$ in Equation~\ref{eq:NuebarSpecCalcMod} with $f_i + \delta f_i$, 
and $n_{\nu}$ with $n_{\nu} + \delta n_{\nu}$.
By substituting the 159 values of $\delta f_i/f_i$ into Equation~\ref{eq:dnn_f}, we 
maintain the implicit correlations between different isotopes. 
And as explained above, the normalization by the thermal power removes the
influence of the $W_{th}$ uncertainty on the $\bar{\nu}_e$ signal, so
that $\sigma_f$ accounts for the contribution from $\delta f_i/f_i$ only. 
To compute the values of the $n_i$ in Equation~\ref{eq:dnn_f}, 
we assume no oscillation (i.e.~short $L$) and an
energy-independent efficiency above the inverse beta-decay threshold.
For the $f_i/F$ we use the same sets of values used to evaluate $\sigma_e$
listed in Table~\ref{tab:reactor_info}. 

\begin{figure}
\begin{center}
\leavevmode
\includegraphics[width=8.6cm]{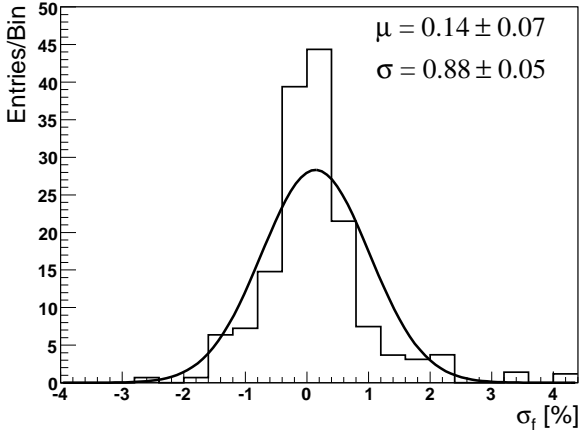}
\end{center}
\caption{\label{fig:dnn} $\sigma_f$ distribution for the fuel comparisons, 
calculated with Eq.~\ref{eq:dnn_f}, weighted by sample burnup and fit to
a Gaussian. The mean of the fit
is identified as the systematic component of $\sigma_f$, while the spread 
represents the statistical uncertainty. The inclusion of a disproportionately
large number of samples from regions that are challenging to simulate
contributes larger tails and a higher mean, leading to more conservative
estimates uncertainties. }
\end{figure}

An example $\sigma_f$ distribution is drawn in
Figure~\ref{fig:dnn} using the fission fractions from~\cite{kamland_prl_2}, 
listed in the ``KL'' row of Table~\ref{tab:reactor_info}.
When drawing the distribution of $\sigma_f$, we weight the value from each 
spent fuel element comparison according to its burnup to account
for the fact that the higher-burnup samples near the center of the reactor
contribute proportionally more to the $\bar{\nu}_e$ signal than the lower-burnup
samples near the core edges. The set of comparisons still includes a disproportionately
large number of fuel elements from regions that are challenging for the simulation;
by including all such samples we ensure a conservative estimate of $\sigma_f$.
These problematic samples are responsible for the relatively large tails and the few
outliers in the distribution. In order to incorporate these samples without
letting them dominate the characterization of the distribution, we fit the distribution to 
a single Gaussian, and take the fit mean and standard deviation as estimates of the systematic 
and statistical components, respectively, of $\sigma_f$.
We repeated this procedure for each of the four choices of the $f_i/F$ listed
in Table~\ref{tab:reactor_info}. As can be seen from these examples, 
$\sigma_f$ is typically about 0.9\%, of which $\sim$0.1\% is correlated between
different reactors.

\begin{table}
\caption{\label{tab:reactor_info} Fission fractions $\frac{f_i}{F}$, $\sigma_e$, and $\sigma_f$ for the three
PVNGS reactors (PV1, PV2, PV3), 
and for all reactors viewed by KamLAND~\cite{kamland_prl_2} (KL), averaged over the periods of operation.}
\scriptsize
\begin{ruledtabular}
\begin{tabular}{lcccccc}
\multirow{2}{0.1\textwidth}{Reactor or  reactor group} & \multicolumn{4}{c}{$\frac{f_i}{F}$ [\%]} & $\sigma_e$ [\%] &$\sigma_f$ [\%] \vspace{0.8mm} \\
& $^{235}$U & $^{238}$U & $^{239}$Pu & $^{241}$Pu &       & syst $\pm$ stat \vspace{0.4mm} \\ 
\hline
PV1    &   58.0    & 7.4       &  29.2      &  5.4       & 0.217 & 0.13 $\pm$ 0.84 \\
PV2    &   54.4    & 7.5       &  31.8      &  6.3       & 0.216 & 0.13 $\pm$ 0.89 \\
PV3    &   57.7    & 7.4       &  29.2      &  5.7       & 0.216 & 0.13 $\pm$ 0.84 \\
KL     &   56.3    & 7.9       &  30.1      &  5.7       & 0.216 & 0.14 $\pm$ 0.88 \\ 
\end{tabular}
\end{ruledtabular}
\end{table}

For an experiment at a single reactor, unless the experimenters and
reactor operators have gone to extremes to validate the core
simulations, the uncertainty in the signal due to the error in the
fission fraction calculations should be taken to be the full
statistical uncertainty on $\sigma_f$ added in quadrature
with the systematic component.  As an example,
using the values listed in Table~\ref{tab:table4} with Type I transit-time UFMs
and incorporating the enthalpy uncertainties (as given in Table~\ref{tab:table3}) 
to obtain the thermal power uncertainty, the resulting uncertainty in
the anti-neutrino yield, ignoring $\sigma_{other}$, is
\begin{eqnarray}
\sigma_{n_\nu} & \approx & \sqrt{ \sigma_W^2 + \sigma_f^2+ \sigma_e^2} \nonumber \\
& \approx & {\scriptstyle \sqrt{\left(0.2^2 + 0.4^2 + 0.15^2 + 0.2^2\right) + \left(0.14^2 + 0.9^2\right) + 0.2^2}} \nonumber \\
& \approx & 1.1\%. \nonumber \\
\end{eqnarray}
If the reactor instead measures the thermal power output at the 2\% level, then this
uncertainty evaluates to 2.2\%. 

For a multi-reactor experiment, some reduction of the statistical components
of $\sigma_W$ and $\sigma_f$ is achieved.
If reactor $k$ contributes a fraction
$p_k$ to the total reactor $\bar{\nu}_e$ signal, then the total random error in the summed 
signal is reduced by a factor of $\sqrt{\sum_k p_k^2}$.
For example, consider the case of an experiment receiving
an approximately equal signal from 2 reactors equipped with
Type I transit-time UFMs. 
The resulting uncertainty in the anti-neutrino yield (again, ignoring $\sigma_{other}$) is
\begin{eqnarray}
\sigma_{n_\nu} & \approx & {\scriptstyle \sqrt{\left(\frac{0.2^2}{2} + 0.4^2 + \frac{0.15^2}{2} + 0.2^2\right) + \left(0.14^2 + \frac{0.9^2}{2}\right) + 0.2^2}} \nonumber \\
& \approx & 0.83\%. \nonumber \\
\end{eqnarray}

As a more complex example, we study the thermal power and fission
fraction uncertainties in the KamLAND experiment.
We use the actual distance and nominal thermal
power of the reactors to compute the fractions $p_k$ they contribute
to the signal~\cite{zelimir_thesis}.
We then allow all random uncertainties to be reduced by a factor 
of $\sqrt{\sum_k p_k^2} \approx \frac{1}{5}$. For the fission fraction uncertainty,
the 0.88\% statistical contribution for a single reactor is reduced to 0.18\%.
However, for KamLAND we must add an additional uncertainty, $\sigma_f^{MC}$, to account
for the fact that to compute the fission fractions
the KamLAND collaboration used a generic reactor simulation 
which was found to agree with detailed simulations, such as those
considered in this paper, at the 1\% level~\cite{KLReactorSignal}.
In this reference, this uncertainty appears to be dominated by random errors, so it may be
appropriate to allow this component to also be reduced by a factor of $\sim$$\frac{1}{5}$.

For the thermal power uncertainty in the KamLAND example, we consider three scenarios:
the standard KamLAND assumption of an across-the-board
2.1\% systematic uncertainty, a slightly more realistic yet still conservative 
assumption that all Japanese reactors are equipped with properly-calibrated Venturi 
flow meters, and an ideal scenario in which all Japanese reactors are equipped with 
ultrasonic flow meters.
The resulting values of $\sigma_{n_\nu}^{\textrm{total}}$ are listed in Table~\ref{tab:KL_total}.
For the standard KamLAND 2.1\% thermal power error and with $\sigma_f^{MC}$ 
taken to be the full 1\%, the contributions of $\sigma_f$ and $\sigma_e$ are negligible,
giving a result consistent with the 2.3\% uncertainty used by the KamLAND collaboration, and
justifying a classification of the KamLAND treatment as ``conservative''.
However, if we make the assumption that Japanese reactors
are equipped with flow meters at least as performant as Venturi flow meters, 
and further take the spread in $\sigma_f^{MC}$
evident in \cite{KLReactorSignal} to indicate that this term is predominantly random in
nature, significant cancellations can be achieved. 
In this case, the systematic uncertainty due to the fouling is the
dominant component; including all other components gives a total uncertainty of 
0.76\%. In an ideal scenario in which all Japanese reactors are equipped with Type I
transit-time UFMs, the total uncertainty would drop to 0.59\%. The dominant component
is still the systematic uncertainty of the flow meter, but the relative contribution of the
other terms is significant. 

\begin{table}
\caption{\label{tab:KL_total}
Signal uncertainties at KamLAND, $\sigma_{\nu}$, obtained by combining $\sigma_{W}$,
in quadrature with $\sigma_f$ and $\sigma_e$.
The calculations were performed assuming various power uncertainties.
The 2.1\% entry corresponds to the value used by the KamLAND collaboration.
}
\scriptsize
\begin{ruledtabular}
\begin{tabular}[t]{ccccccc} 
$\sigma_{W}^{\mathrm{single}}$[\%] & $\sigma_{W}$[\%] & $\sigma_{f}$[\%] & $\sigma_f^{MC}$[\%] & $\sigma_e$[\%] & $\sigma_{n_\nu}$[\%] \\

syst$\pm$stat   & syst$\pm$stat   & syst$\pm$stat   &     &    &       \\\hline
2.10$\pm$0.00   & 2.10$\pm$0.00   & 0.14$\pm$0.18    & 1.0 & 0.216   &  2.3 \\ 
0.60$\pm$1.40   & 0.60$\pm$0.26   & 0.14$\pm$0.18    & 0.2 & 0.216   &  0.76 \\    
0.45$\pm$0.25   & 0.45$\pm$0.05   & 0.14$\pm$0.18    & 0.2 & 0.216   &  0.59 \\ 
\end{tabular}
\end{ruledtabular}
\end{table}

The actual situation in the Japanese
reactor industry is probably somewhere in between the all-Venturi
case and the ideal case of all UFMs. Many reactors may
already be equipped with UFMs, but some may not, and those that are may
simply use them to calibrate their Venturi meters.
However, the total uncertainty, $\sigma_{n_\nu}$, 
does not change much between these two cases.
Thus for KamLAND, a reduced error of 0.6-0.8\% may be more appropriate. 
A better understanding of the nature of the Japanese instrumentation
and reactor simulations may result in yet smaller overall uncertainties for KamLAND. 

\section{Conclusion}\label{conclusion}

The uncertainty in the estimated anti-neutrino interaction rate 
at reactor experiments has contributions from
the thermal power and the fission calculation uncertainties.
The dominant contributions to the thermal power uncertainty come from the
enthalpy rise and mass flow rate in the steam generator (PWR) or reactor vessel (BWR).
The former has a systematic uncertainty of 0.2\% and a random uncertainty of typically 0.15\%.
The uncertainty in the mass flow rate depends on the instrumentation used to perform
the measurement. Traditional Venturi flow meters exhibit significant fluctuations of 1.4\%
and a systematic uncertainty of 0.6\% due to fouling. More precise ultrasonic flow meters
have been evaluated to have systematic uncertainties below 0.5\% and typically show
random fluctuations below this level.

We estimated, for the first time to our knowledge, the contribution to the $\bar{\nu}_e$
signal uncertainty from errors in the fission rate calculations. These
errors were extracted from comparisons of heavy element concentration measurements in
spent fuel elements with calculations of those concentrations by the same
simulations that reactor $\bar{\nu}_e$ experiments rely on to obtain the
fraction of fissions from each of four main heavy isotopes. Error propagation
accounting for correlations between the four isotopes yielded a 0.1\% systematic
uncertainty on the $\bar{\nu}_e$ signal rate, and a 0.9\% random uncertainty. We also
estimated a 0.2\% contribution due to the uncertainty in the measured energy release per
fission of each isotope.

We demonstrated how these errors can be combined for typical reactor experiment
configurations, and highlighted situations in which cancellations occur between
different reactor sites. With this methodology, we found that these contributions
to the signal rate uncertainty in the KamLAND experiment may be reduced from 
2.3\% to 0.76\%, or even 0.59\%, pending better understanding of reactor instrumentation.
As we move deeper into the phase of precision reactor neutrino oscillation experiments,
and in particular multi-reactor $\theta_{13}$ experiments, such sub-percent understanding
of these errors and their systematic and random components will become more
and more important. As reactor engineering, flow rate metrology, and core modeling
continue to improve, further reductions may be achievable in these uncertainties.
Toward this end, closer relationships between neutrino scientists and reactor operators 
should be encouraged.

\begin{acknowledgments}
We are grateful to Harry Miley for providing some of the references
used here and to Palo Verde Nuclear Generating Station for
providing data used in the analysis.
We thank Petr Vogel and Patrick Decowski for their careful reading and useful comments. 
We would also like to thank Akira Sebe and Kazuhiro Terao for their help in
translating the Japanese references.
This work was supported, in part, by U.S. National Science Foundation grant no.~PHY-0758118, 
U.S. Department of Energy contract no.~DE-AC02-05CH1123 and grant nos.~DE-FG02-01ER41166 
and~DE-FG02-04ER41295.
\end{acknowledgments}

\end{document}